\documentstyle[aps,epsf,prl,twocolumn]{revtex}
\begin{document}
\title{Magnetic Resonance Force Microscopy Measurement of Entangled Spin States
}
 \author{G.P. Berman,$\!^1$ F. Borgonovi,$\!^{1,2}$ G. Chapline,$\!^{1,3}$ P.C. Hammel$^4$, and V.I. Tsifrinovich$^5$}
 \address{$^1$Theoretical Division and CNLS,
 Los Alamos National Laboratory, Los Alamos, NM 87545}
 \address{$^2$Dipartimento di Matematica e Fisica, Universit\`a Cattolica,
 via Musei 41 , 25121 Brescia, Italy, and I.N.F.M., Gruppo Collegato
 di Brescia, Italy, and I.N.F.N., sezione di Pavia , Italy}
 \address{$^3$Lawrence Livermore National Laboratory, Livermore, CA 94551}
 \address{$^4$Condensed Matter and Thermal Physics, Los Alamos National Laboratory,
MS K764, Los Alamos NM 87545}
 \address{$^5$IDS Department, Polytechnic University,
 Six Metrotech Center, Brooklyn NY 11201}
\maketitle
%
%
{We simulate magnetic resonance force microscopy measurements of
an entangled spin state. One of the entangled spins drives the
resonant cantilever vibrations, while the other remote spin does
not interact directly with the quasiclassical cantilever. The
Schr\"odinger cat state of the cantilever reveals two possible
outcomes of the measurement for both entangled spins. }
\\[1ex]

Magnetic resonance force microscopy (MRFM) proposed a decade ago
\cite{1} is now approaching its ultimate goal: single spin
detection \cite{2}. The following question arises: To what extent
can the MRFM be used for quantum measurement of spin states? The
particular problem considered in this paper is the MRFM
measurement of entangled spin states using a cyclic adiabatic inversion of the spin, which drives the resonant vibrations of the cantilever. We discuss the possibility
of determining the state of a remote spin which is entangled with
the spin interacting with the MRFM measurement apparatus. 
 \begin{figure}
 \epsfxsize 7cm
 \epsfbox{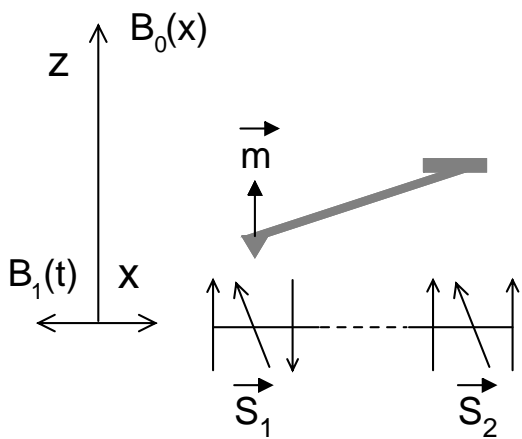}
 \narrowtext
 \caption{
Diagram of an MRFM measurement of an entangled state of two spins
in a chain of spins.  $\vec{S}_1$ is a single measured spin which
changes its direction under the action of the alternating magnetic
field $B_1$. ${\vec S}_2$ is the remote spin entangled with the
spin $\vec{S}_1$. $\vec{m}$ is the magnetic moment of a
ferromagnetic particle which is attached to the cantilever tip.
The magnetic force on the cantilever tip (attractive or
repulsive) depends on the direction of the spin $\vec{S}_1$.
$B_0$ is the permanent magnetic field which points in the
positive $z$-direction.} \label{canti}
\end{figure}
As an
example, we consider a typical entangled state for two spatially
separated spins,
$$
(1/\sqrt{2})(|\uparrow\uparrow\rangle
+|\downarrow\downarrow\rangle).\eqno(1)
$$
According to the conventional point of view, by measuring the
left spin in the state $|\uparrow\rangle$ (or
$|\downarrow\rangle$) we automatically collapse a remote right
spin into the same state $|\uparrow\rangle$ (or
$|\downarrow\rangle$). In the process of MRFM measurement (see
Fig. 1) the direction of the measured spin, $\vec{S}_1$, changes
periodically with the period of the cantilever vibrations. Thus,
it is not clear what the direction of the entangled remote spin,
$\vec{S}_2$, will be after the MRFM measurement.

To study this problem, we simulated the quantum dynamics of this
spin-cantilever system assuming that the measuring spin is
initially entangled with the remote spin. The remote spin is not
subjected to the action of the MRFM apparatus.

The dimensionless quantum Hamiltonian of the spin-cantilever
system in the rotating reference frame is \cite{3},
$$
{\cal H}=(p^2_z+z^2)/2+\dot\varphi S_{z1}-\epsilon S_{x1}-2\eta zS_{z1}.\eqno(2)
$$
Here $p_z$ and $z$ are the dimensionless momentum and coordinate
of the cantilever tip; ${\vec S}_{1}$ is the ``first'' measured
spin;  $\epsilon=\epsilon(\tau)$ is the dimensionless amplitude
of the radio-frequency ({\it rf}) field (where $\tau=\omega_ct$
is the dimensionless time and $\omega_c$ is the cantilever
frequency); $\eta$ is the dimensionless constant of interaction
between the cantilever and the spin, which is proportional to the
magnetic field gradient produced by the ferromagnetic particle on
the cantilever tip. The phase of the {\it rf} field is taken in
the form $(\omega t+\varphi(t))$, where $\omega$ is chosen equal
to the Larmor frequency of the spin: $\omega=\omega_L$. The time
derivative, $\dot\varphi$, changes periodically with the
frequency of the cantilever vibrations, $\omega_c$. In our
notation, the dimensionless frequency of the cantilever
vibrations is one unit. Thus, the dimensionless period of the
cantilever vibrations is $2\pi$. The periodic oscillation of
$\dot\varphi$ provides a cyclic adiabatic inversion of the spin
\cite{4}, which drives the resonant vibrations of the cantilever.

The dimensionless wave function of the whole system, including
the second entangled spin, can be written in the
$z$-representation as,
$$
\Psi(z,S_{z1},S_{z2},\tau)=
u_{\uparrow\uparrow}(z,\tau)|\uparrow\uparrow\rangle+
u_{\uparrow\downarrow}(z,\tau)|\uparrow\downarrow\rangle+\eqno(3)
$$
$$
u_{\downarrow\uparrow}(z,\tau)|\downarrow\uparrow\rangle+
u_{\downarrow\downarrow}(z,\tau)|\downarrow\downarrow\rangle.
$$
Substituting (3) into the Schr\"odinger equation, we derive four
coupled equations for the functions $u(z,\tau)$,
$$
2i\dot u_{\uparrow\uparrow}=(p^2+z^2+\dot\varphi-2\eta
z)u_{\uparrow\uparrow}-\epsilon u_{\downarrow\uparrow},\eqno(4)
$$
$$
2i\dot u_{\downarrow\uparrow}=(p^2+z^2-\dot\varphi+2\eta
z)u_{\downarrow\uparrow}-\epsilon u_{\uparrow\uparrow},
$$
$$
2i\dot u_{\uparrow\downarrow}=(p^2+z^2+\dot\varphi-2\eta
z)u_{\uparrow\downarrow}-\epsilon u_{\downarrow\downarrow},
$$
$$
2i\dot u_{\downarrow\downarrow}=(p^2+z^2-\dot\varphi+2\eta
z)u_{\downarrow\downarrow}-\epsilon u_{\uparrow\downarrow}.
$$
This system of equations splits into two independent sets of
equations. 
\begin{figure}
\epsfxsize 8cm
\epsfbox{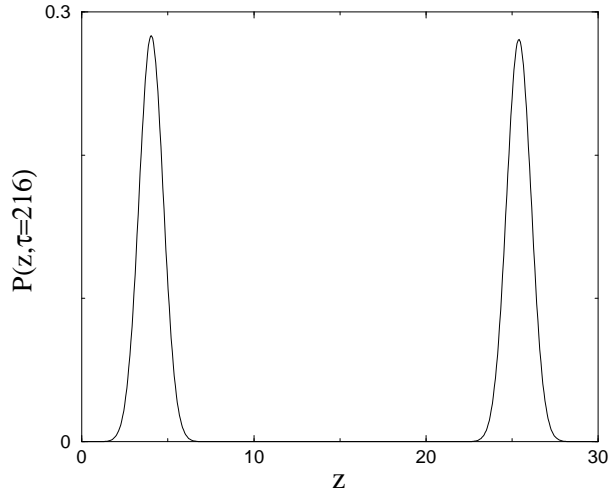}
\narrowtext
\caption{
The probability distribution, $P(z)$, at $\tau=216$.}
\label{canti}
\end{figure}
The initial condition is assumed to be a product of
the coherent quasiclassical wave function of the cantilever and
the entangled state of the two spins,
$$
\Psi(z,S_{z1},S_{z2},0)=(1/\sqrt{2})u_\alpha(z)
(|\uparrow\uparrow\rangle+|\downarrow\downarrow\rangle),\eqno(5)
$$
$$
u_\alpha(z)=\pi^{-1/4}\exp[-(z-\sqrt{2}\alpha)^2/2],
$$
where $z_0\equiv \sqrt{2}\alpha$ is the average initial
coordinate of the cantilever. For numerical simulations we used
the following values of parameters,
$$
\alpha=-10\sqrt{2},~\eta=0.3,
$$
and the following time dependences for $\epsilon(\tau)$ and
$\dot\varphi(\tau)$,
$$
\epsilon=20\tau,~\tau\le 20;~\epsilon=400,~\tau>20,\eqno(6)
$$
$$
\dot\varphi=-600+30\tau,~\tau\le 20;~\dot\varphi=1000\sin(\tau-20),~\tau>20.
$$
The chosen value of $\alpha$ corresponds to the quasiclassical
state of the cantilever with the number of excitations
$n=|\alpha|^2=200$. The chosen value of $\eta$ corresponds to the
existing MRFM experiments with electron spins \cite{2}. The time
dependences (6) provide the conditions for the cyclic adiabatic
inversion of the spin. Fig. 2 shows the typical probability
distribution of the cantilever position,
$$
P(z)=|u_{\uparrow\uparrow}|^2+|u_{\uparrow\downarrow}|^2
+|u_{\downarrow\uparrow}|^2+|u_{\downarrow\downarrow}|^2.\eqno(7)
$$
One can see that the probability distribution, $P(z)$, describes
a Schr\"odinger cat state of the cantilever with two
approximately equal peaks. When these two peaks are clearly
separated, the total wave function can be represented as a  sum
of two terms corresponding to the ``left'' and the ``right''
peaks in the probability distribution,
$$
\Psi(z,S_{z1},S_{z2},\tau) =\Psi_a(z,S_{z1},S_{z2},\tau)
+\Psi_b(z,S_{z1},S_{z2},\tau).\eqno(8)
$$
Our numerical analysis shows that each term, $\Psi_a$ and
$\Psi_b$, can be approximately decomposed into a direct product
of the cantilever and spin wave functions,
$$
\Psi_a=u_a(z,\tau)\chi_a(S_{z1},\tau)|\uparrow\rangle_2,
~\Psi_b=u_b(z,\tau)\chi_b(S_{z1},\tau)|\downarrow\rangle_2.\eqno(9)
$$
This decomposition is possible because the complex function $u_{\uparrow\uparrow}(z,\tau)$ is proportional to $u_{\downarrow\uparrow}(z,\tau)$, and  the complex function $u_{\uparrow\downarrow}(z,\tau)$ is proportional to $u_{\downarrow\downarrow}(z,\tau)$. Such proportionality can be seen
in Fig. \ref{psi}, where we plot the corresponding wave functions
at the same time as in Fig. 2, with suitable numerical coefficients.

\begin{figure}
  \epsfxsize 8cm \epsfbox{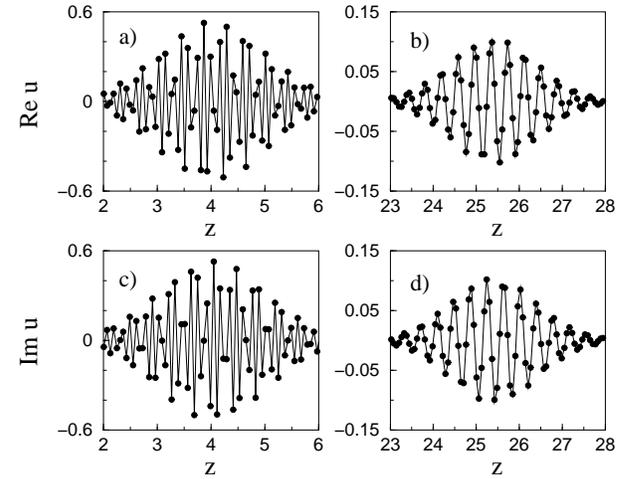} \narrowtext \caption{ Upper
  boxes: the real part of wave functions; lower boxes: the
  imaginary part of wave functions, at $\tau = 216$.
  a)$Re(u_{\uparrow\uparrow})$: solid line;
  $Re(-5u_{\downarrow\uparrow})$: filled circles.
  b) $Re(u_{\uparrow\downarrow})$: solid line;
  $Re(5u_{\downarrow\downarrow})$: filled circles.
  c) $Im(u_{\uparrow\uparrow})$: solid line; circles
  $Im(-5u_{\downarrow\uparrow})$filled circles.
  d) $Im(u_{\uparrow\downarrow})$: solid line;
  $Im(5u_{\downarrow\downarrow})$: filled circles.} \label{psi}
\end{figure}

The spin wave function, $\chi_a(S_{z1},\tau)$, describes the
dynamics of the first  spin with its average, $\langle\chi_a|\vec
S|\chi_a\rangle$, pointing approximately in the direction of the
effective magnetic field, ${\vec B}_{\rm eff}$, in the rotating
frame,
$$
\vec B_{\rm eff}=(\epsilon,0,-\dot\varphi).\eqno(10)
$$
(We neglect here the nonlinear term $2\eta z$ whose contribution
to the effective field is small.) 
\begin{figure}
\epsfxsize 8cm
\epsfbox{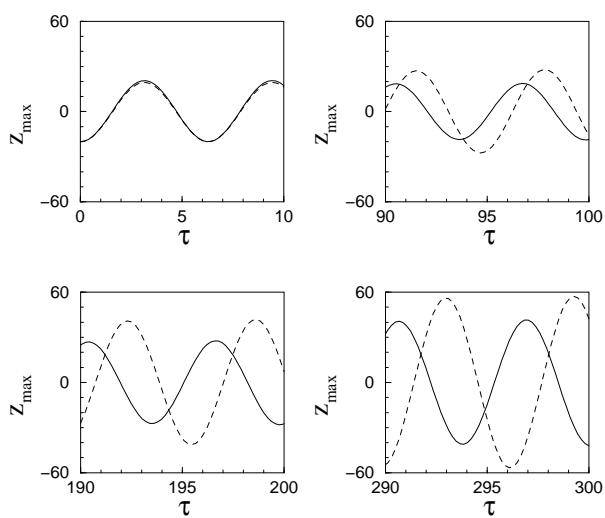}
\narrowtext
\caption{The positions, $Z_{max}(\tau)$, of two peaks of the
Schr\"odinger cat state as a function of time.}
\label{canti}
\end{figure}
The spin function,
$\chi_b(S_{z1},\tau)$, describes the dynamics of the first spin
with its average pointing in the direction opposite to the
direction of $\vec B_{\rm eff}$. 
As the amplitude of the
cantilever vibrations increases, the phase difference between the
oscillations of the two peaks, $|u_a(z,\tau)|^2$ and
$|u_b(z,\tau)|^2$, approaches $\pi$. (See Fig. 4.) (To reach the
phase difference of $\pi$, a long time of numerical simulations
is required. That is why we restricted the simulation time for
the results presented in Fig. 4.)
\begin{figure}
\epsfxsize 8cm
\epsfbox{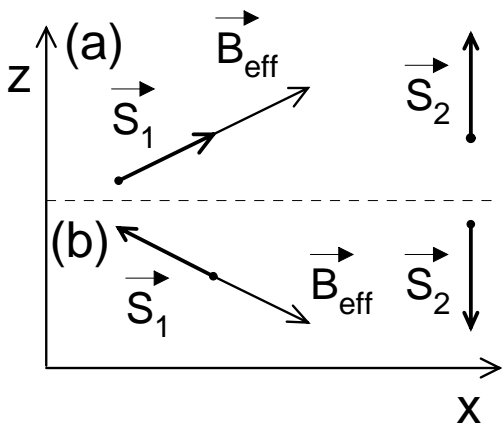}
\narrowtext
\caption{Two outcomes of the MRFM measurement of the state of two
entangled spins.  (a) The measured spin, $\vec {S_1}$, points
along the direction of the effective magnetic field, and the
remote spin, $\vec {S_2}$, points ``up'' (in the positive
$z$-direction). (b) The measured spin, $\vec {S_1}$, points in
the direction opposite to the effective magnetic field, and the
remote spin, $\vec {S_2}$, points ``down'' (in the negative
$z$-direction).}
\label{canti}
\end{figure}
 In realistic experimental conditions, the Schr\"odinger cat
 state quickly collapses due to the interaction with the
 environment \cite{5}. In this case, the two peaks of the
 probability distribution describe two possible trajectories of
 the spin-cantilever system.
In one of these trajectories the first (measured) spin is pointed
along the direction of the effective magnetic field while  the
second (remote) spin is pointed ``up'' (in the positive
$z$-direction); the other trajectory corresponds to the opposite
situation in which the orientation of both spins is reversed---the
first (measured) spin is antiparallel to the effective magnetic
field, and the second (remote) spin is pointed ``down'' (in the
negative $z$-direction). The phase difference between the
corresponding oscillations of the cantilever approaches $\pi$
with increasing cantilever vibration amplitude.

In summary, we have studied the outcome of the MRFM measurement
of the entangled spin state,
$(1/\sqrt{2})(|\uparrow\uparrow\rangle
+|\downarrow\downarrow\rangle)$. Our numerical simulations reveal
two possible outcomes shown schematically in Fig. 5: (a) The
first (measured) spin points along the effective magnetic field
in the rotating frame, and the second (remote) spin points in the
positive $z$-direction. (b) The first (measured) spin points
opposite to the direction of the effective magnetic field in the
rotating frame, and the second (remote) spin points in the
negative $z$-direction. Thus, the collapse of the measured spin
along (or opposite) the direction of the rotating effective
magnetic field leads to the collapse of the remote spin in the
positive (or negative) $z$-direction. These two outcomes
correspond to two phases  of the cantilever vibrations which
differ by $\pi$.

This work  was supported by the Department of Energy under contract
W-7405-ENG-36 and DOE Office of Basic Energy Sciences. The work of
GPB, PCH and VIT was partly supported by the National
Security Agency (NSA) and by the Advanced Research and
Development Activity (ARDA).

\end{document}